\documentclass[12pt,letterpaper]{article}
\usepackage[T1]{fontenc}
\usepackage{lmodern}
\usepackage{microtype}
\usepackage[letterpaper,margin=1in]{geometry}
\usepackage{setspace}
\usepackage{amsmath,amssymb,amsthm,mathtools,bm}
\usepackage{graphicx,booktabs,tabularx,enumitem}
\usepackage{tikz}
\usetikzlibrary{arrows.meta,calc,positioning}
\usepackage[round,authoryear,sort]{natbib}
\setcitestyle{aysep={},citesep={;}}
\usepackage[hidelinks]{hyperref}
\newtheorem{theorem}{Theorem}[section]

\newtheorem{corollary}[theorem]{Corollary}

\theoremstyle{definition}
\newtheorem{definition}[theorem]{Definition}
\newtheorem{example}[theorem]{Example}

\newcommand{\ip}[2]{\langle #1,#2\rangle}
\newcommand{\norm}[1]{\lVert #1\rVert}
\newcommand{\R}{\mathbb{R}}
\newcommand{\one}{\mathbf{1}}
\newcommand{\E}{\mathbb{E}}
\newcommand{\Prob}{\mathbb{P}}
\newcommand{\dd}{\mathrm{d}}
\DeclareMathOperator{\diag}{diag}
\DeclareMathOperator{\tr}{tr}
\DeclareMathOperator{\rank}{rank}
\DeclareMathOperator{\im}{im}
\DeclareMathOperator{\logit}{logit}
\DeclareMathOperator{\Var}{Var}
\DeclareMathOperator*{\argmin}{arg\,min}

\tikzset{
  obj/.style={circle,draw,minimum size=7mm,inner sep=0pt},
  win/.style={-{Latex[length=2mm]},thick},
  missing/.style={dashed,gray},
  elab/.style={midway,fill=white,inner sep=1pt,font=\scriptsize}
}

\hypersetup{
  pdftitle={Likelihood-Free Significant Rankability Test through Hodge Projection},
  pdfauthor={Hengrui Luo and Yufeng Liu}
}
\title{\bfseries Likelihood-Free Significant Rankability Test through Hodge Projection}
\author{Hengrui Luo\thanks{Email: hrluo@rice.edu}\\
  \small Department of Statistics, Rice University\\
  \small Computational Research Division, Lawrence Berkeley National Laboratory
  \\[0.7em]
  Yufeng Liu\thanks{Email: yufliu@umich.edu}\\
  \small Department of Statistics, University of Michigan, Ann Arbor}
\date{}
\begin{document}
\maketitle

\begin{abstract}
Pairwise comparison data, such as human preference judgments between AI systems, often exhibit a score-difference structure in which objects can be approximately ranked so that observed outcomes reflect differences in latent scores. We introduce SigRank, a likelihood-free test for whether such structure is present in an observed flow of pairwise comparisons that may be incomplete, weighted, or contain repeated edges. SigRank projects the flow onto the gradient space and calibrates the normalized projection energy under the null of independent fair edge signs, conditional on the observed graph, weights, and magnitudes. Unlike lack-of-fit tests for a ranking model or tests of ordinal acyclicity, it detects an unusually large score-difference component without assuming a response link. HodgeRank supplies the projection estimator. Our main contribution is a conditional inferential calibration of the Hodge projection. We establish exact finite-sample validity, prediction and score-error bounds, dense-graph power, a scale-invariant conditional quadratic-form limit, and sensitivity formulas for comparison reversals. On complete unweighted binary tournaments, the statistic is an affine transform of the circular-triad count. In matched Chatbot Arena analyses,  SigRank and Bradley--Terry agree for LLaMA but give contrasting evidence for Pythia, illustrating why a fitted score contrast can warrant caution about broader ranking claims. Non-rejection leaves the presence of ranking structure unresolved and does not establish its absence.
\end{abstract}
\noindent\emph{Keywords: Pairwise ranking, Nonparametric hypothesis test, Hodge decomposition}
\setlength{\abovedisplayskip}{8pt plus 2pt minus 2pt}
\setlength{\belowdisplayskip}{8pt plus 2pt minus 2pt}
\setlength{\abovedisplayshortskip}{4pt plus 2pt minus 1pt}
\setlength{\belowdisplayshortskip}{6pt plus 2pt minus 1pt}
\section{Introduction}

Pairwise-comparison data arise naturally in many settings, including consumer preference studies, sports tournaments, and artificial intelligence (AI) model evaluations, and are often incomplete and unevenly replicated \citep{bhattacharya2026inferences}. %
The statistical challenge is that a fitted ranking always exists on a connected comparison graph, even when the observed directions are unrelated to any common preference order. We therefore ask whether the observed flow contains more additive score-difference alignment than random directions would produce, conditional on the graph, weights, and magnitudes.  %
Let $N_{ij}$ and $N_{ji}$ denote the number of wins of unit $i$ over $j$ and unit $j$ over $i$, respectively, and $y_{ij}\in [-1,1]$ summarizes the net outcome for pair $(i,j)$, where the units are  specified before testing. If units are treated as nodes and individual judgments are treated as  edges between nodes, the null randomizes each primitive sign of $y_{ij}$   independently and %
imposes different sign-symmetry and independence assumptions.

A globally rankable observed edge flow has the form $y_{ij}=\theta_i-\theta_j$, where $\theta_i$ and $\theta_j$ are latent scores for objects $i$ and $j$. This geometry underlies score-based ranking methods \citep{UematsuLee2015,UematsuLee2017}. The Bradley-Terry (BT) model \citep{BradleyTerry1952} uses these score differences on a logistic response scale. Writing $i\succ j$ for a win of $i$ over $j$ in pair $(i,j)$, the BT model assumes
\begin{equation}\label{eq:logistic_link}
\logit\Pr(i\succ j)=\theta_i-\theta_j,
\qquad
\Pr(i\succ j)=\sigma(\theta_i-\theta_j),
\end{equation}
where $\sigma(t)=(1+e^{-t})^{-1}$ and $\logit(p)=\log\{p/(1-p)\}$ for $0<p<1$. Thurstone models instead use a Gaussian-threshold scale \citep{thurstone1927law}.  Both families assume a specific link function and assess fit or perform inference within that assumed model \citep{makur2025minimax,SinghDavidov2026}. By contrast, our proposed method SigRank does not assume or test either link function. It detects a score-difference component in the chosen observed edge flow without assuming any parametric response model. This makes it a diagnostic when model specification is uncertain. For example, repeated judgments may be dependent or unequally variable, which can miscalibrate an independent-binomial test even when every pair is fair on average. The conditional sign approach tolerates such magnitude variation only when the chosen edge signs remain jointly independent and fair after conditioning. Thus a weaker response-model assumption does not imply validity under arbitrary dependence.

A complementary statistical literature studies what can be estimated once a ranking target or response model has been specified. Under Bradley--Terry--Luce (BTL) and Thurstone-type models, comparison-graph topology controls score-estimation error \citep{ShahEtAl2016}, and uncertainty quantification is available even on sparse  graphs \citep{GaoShenZhang2023}. For cardinal outcomes, graph-based least-squares theory likewise establishes estimation and ranking properties \citep{SinghIliopoulosDavidov2025}. Model-robust methods can recover top-$k$ or full rankings without imposing the BTL likelihood \citep{ShahWainwright2018}, while sensitivity analyses question the stability of fitted rankings \citep{huang2026dropping}. These studies primarily concern estimation or recovery of a ranking, or  uncertainty quantification for its parameters. In related work, cyclicity and lack-of-fit tests ask whether deviations from a posited order are too large \citep{hoffman2022methods,regenwetter2010testing,makur2025minimax,SinghDavidov2024}. SigRank asks a different, preceding question: is the observed score-difference projection unusually strong relative to random directions? A statistically detectable component needs not explain most of the edge variation, and its detection does not validate a particular response link.

SigRank is based on a classical geometric foundation.  HodgeRank projects a comparison flow onto the gradient space and decomposes the residual into curl and harmonic components \citep{Jiang2011,oliveira2024hodgerank}.  Given the same edge flow, weights, and norm, the SigRank effect size is one minus the corresponding normalized Hodge residual energy.  HodgeRank describes the projection, whereas SigRank assigns it a conditional significance level. The use of independent signs given magnitudes recalls the signed-rank principle \citep{Wilcoxon1945} and differs from binomial or Gaussian model-based calibration. SigClust likewise asks whether apparent structure exceeds a null benchmark, using a two-cluster index under a fitted single-Gaussian model \citep{Liu2008}. SigRank instead fixes the Hodge projection and tests independent fair edge orientations. The analogy is conceptual: neither the null hypothesis nor the structural target is the same.

A classical special case connects SigRank to the circular-triad literature \citep{KendallSmith1940,Jiang2011}:  for one non-tied binary comparison per pair on a complete unweighted graph, the normalized gradient projection energy (Definition~\ref{def:sigrank}) is an affine decreasing function of the number of circular triads, so a fair-sign orbit test has the same rejection ordering as a circular-triad test (See Supplementary Proposition~S2.1).  This relation,  %
consistent with the connection between Hodge curl \citep{KendallSmith1940} and the Kendall-Babington Smith consistency index \citep{Jiang2011}, is a useful check rather than a general reduction: it does not extend in general to incomplete graphs, heterogeneous magnitudes or weights, or repeated records treated as separate edges.

Deterministic combinatorial rankability measures quantify how close a dominance graph is to a uniquely rankable tournament \citep{AndersonChartierLangville2019,AndersonEtAl2021Weighted}.  Recent graphical paired-comparison work estimates or selects cyclic components and tests lack of fit to linear stochastic transitivity \citep{SinghDavidov2024,SinghDavidov2026}.  Those lack-of-fit procedures assume a score-difference structure under the null and reject for excess cyclic residuals. SigRank instead tests independent fair directions conditional on magnitudes and rejects for unusually strong gradient alignment. Therefore, SigRank can test single observation per pair (with potential ties), while existing methods like \citet{SinghDavidov2024,SinghDavidov2026} and \citet{SinghIliopoulosDavidov2025} will need additional information for their lack-of-fit procedures, such as the noise variance or a prespecified set of potentially cyclicity under complete-binary parametric models.
Recent inference for circular-triad counts in inhomogeneous random tournaments also uses a different stochastic model and is restricted to complete binary tournaments \citep{ChatterjeeBhattacharya2026}.

The rankability raise a complementary point: detection of a gradient component does not validate a parametric link or guarantee a good global fit.  To our knowledge, %
no prior work conditions simultaneously on an arbitrary observed comparison graph, beyond the complete-binary special case noted above, to test for a large score-difference projection. In short, SigRank provides finite-sample likelihood-free inference for the presence of a score-difference component under a conditional fair-sign null, leaving the distribution of margin sizes unspecified.

The proposed SigRank method has three parts. First, weighted Hodge projection summarizes the observed flow by fitted score differences and their explained fraction of edge energy. Second, the observed comparison graph and weights fix the score-difference subspace, which determines what evidence can accumulate across pairs. Third, the conditional sign null (in \eqref{eq:null}) calibrates that evidence while holding the graph, weights, and magnitudes fixed. ``Likelihood-free'' means that no response link is specified.

Our main contribution is conditional inference for HodgeRank projection, namely SigRank. Exact randomization validity supports testing without a response likelihood, prediction and score-error bounds separate reliable edge fitting from accurate object scores, and the dense-graph power result identifies a regime of reliable detection. The conditional quadratic-form limit provides a scalable approximation when edge and spectral influences are sufficiently diffuse. Sign-reversal bounds quantify how label errors can change the evidence. Together, these results connect the test construction to its assumptions, operating regime, and practical limitations. %
Chatbot Arena analyses provide detailed comparisons for the LLaMA and Pythia groups. Significance supplies conditional evidence for a score-difference component, while non-rejection leaves this question unresolved and warrants caution when applying likelihood-based BT models. %

The remainder of this article is organized as follows: Section~\ref{sec:matrix-sigrank} develops SigRank and its illustrative examples. Section~\ref{sec:theory} establishes its statistical properties. Section~\ref{sec:simulation} examines calibration, power, and discrimination under within-pair dependence and heterogeneous dispersion. Section~\ref{sec:llm-arena} presents the matched Chatbot Arena analyses. %

\section{SigRank\label{sec:matrix-sigrank}}
We first review the Hodge projection, then define its normalized effect size and conditional sign calibration. The examples separate a large descriptive projection from statistically calibrated evidence. Proofs of the projection and finite-sample validity are in Supplementary Section~S1.

Before the formal development, we state the test's hypotheses for SigRank in plain terms:
\begin{itemize}
  \item[$H_0$:] Conditional on the chosen edges, their weights, and their absolute margins, the edge signs are independent fair coin flips.
  \item[$H_1$:] The directions exhibit unusually strong alignment with the gradient subspace, that is the observed flow has more score-difference structure than independent fair signs would produce given the same magnitudes. %
\end{itemize}
A small $p$-value means the data contain more score-difference structure than chance alone can explain. A large $p$-value means the observed pattern is compatible with random directions. The formal version of $H_0$ is given in~\eqref{eq:null}.
\subsection{Finite-sample Hodge theory and Hodge projections}
\label{sec:finite_sample_hodge}
Fix the edge unit declared before analysis. Let \(G=(V,E)\), \(V=\{1,\ldots,n\}\), be a connected comparison graph with \(m=|E|\) edges and an arbitrary reference orientation. Let \(y=(y_e:e\in E)\in\mathbb R^m\) be the antisymmetric observed flow. Binary records have \(y_e\in\{-1,+1\}\). Aggregated records use the margin defined in the introduction. Let \(B\in\R^{m\times n}\) be the oriented incidence matrix, so \((B\theta)_{(i,j)}=\theta_i-\theta_j\), and let \(\one\in\R^n\) denote the all-ones vector. With \(W=\diag(w_e:e\in E)\), \(w_e>0\), define \(\ip{u}{v}_W=u^\top Wv\) and \(\norm{u}_W^2=\ip{u}{u}_W\).

Let \(C^0=\mathbb R^V\) be the space of vertex functions, interpreted here as ranking scores \(\theta_i\), and let \(C^1=\mathbb R^E\) be the space of edge flows, interpreted as observed pairwise margins \(y_{ij}\).  The coboundary \(\dd_0:C^0\to C^1\) is given by \(\dd_0=B\), so \((\dd_0\theta)_{ij}=\theta_i-\theta_j\). Thus score-based ranking is precisely the class of gradient  flows on edges.

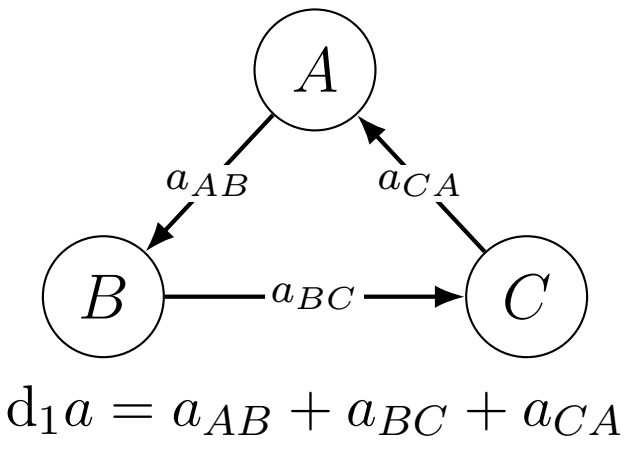
\begin{figure}[h]
\centering
\begin{tikzpicture}[
  obj/.style={circle,draw,minimum size=8mm,inner sep=0pt},
  flow/.style={-{Latex[length=2mm]},thick},
  tri/.style={fill=gray!15,draw=gray!40},
  lab/.style={midway,fill=white,inner sep=1pt,font=\scriptsize},
  scale=1.0
]
  \node[obj] (A) at (0,1.5) {$A$};
  \node[obj] (B) at (-1.4,0) {$B$};
  \node[obj] (C) at (1.4,0) {$C$};

  \draw[flow] (A) -- node[lab] {$a_{AB}$} (B);
  \draw[flow] (B) -- node[lab] {$a_{BC}$} (C);
  \draw[flow] (C) -- node[lab] {$a_{CA}$} (A);

  \node[align=center,font=\small] at (0,-0.75)
  {$\dd_1a=a_{AB}+a_{BC}+a_{CA}$};
\end{tikzpicture}
\caption{\label{fig:incompatible_triangle}A score-difference flow \(a_{ij}=\theta_i-\theta_j\) has zero circulation around every triangle.}
\end{figure}

With the Euclidean inner product on \(C^0\) and the \(W\)-weighted inner product on \(C^1\), the adjoint of \(\dd_0\) is
\[
        \delta_0:C^1\to C^0,
        \qquad
        \delta_0 a=B^\top W a .
\]
Therefore $ L_W=\delta_0\dd_0=B^\top WB $ is the weighted graph Laplacian.  In ranking terms, \(\delta_0 a\) is the weighted net imbalance at each object: positive values indicate net outgoing preference flow, negative values indicate net incoming preference flow.  The gradient subspace \(\im(B)\) is the class of globally orderable observed-edge flows as a chain of consistent directed arrows in the comparison graph. Define the centered weighted least-squares score
\begin{equation}\label{eq:theta-ls}
        \widehat\theta
        =\argmin_{\theta:\,\one^\top\theta=0}
        \norm{y-B\theta}_W^2.
\end{equation}
The score \(\hat\theta\) solves the following linear equation associated with the comparison graph
\begin{align}
\label{Jiang_LS}
        L_W\hat\theta=B^\top Wy,
\end{align}
so it is the score function whose gradient flow best explains the observed pairwise margins.%

If a two-dimensional simplicial complex \(K\) is placed on the graph, let \(F\) denote the set of oriented triangles and let \(C^2=\mathbb R^F\). The triangle coboundary \(\dd_1:C^1\to C^2\) is \((\dd_1 a)_{ijk}=a_{ij}+a_{jk}+a_{ki}\), which measures local cyclic inconsistency \citep{Jiang2011}. For a score-difference flow \(a=\dd_0\theta\), the terms telescope, so \((\dd_1\dd_0\theta)_{ijk}=0\). Hence every globally rankable component has zero triangle circulation.  Any nonzero value of \(\dd_1 y\) indicates a cyclic or matchup-specific component that cannot be represented by one scalar ranking score.  Let \(\delta_1\) be the adjoint of \(\dd_1\) under the chosen triangle weights. The Hodge decomposition separates an observed comparison flow into a gradient ranking part and residual cyclic parts.

\begin{theorem}[Weighted Hodge projection]\label{thm:hodge-projection}
Assume that \(G\) is connected. For every edge flow \(y\in C^1\), there is a unique decomposition
\begin{equation}\label{eq:hodge-basic}
        y = \dd_0\hat\theta + r,
        \qquad
        \dd_0\hat\theta\in\im(\dd_0),
        \qquad
        r\in\ker(\delta_0),
\end{equation}
with \(\one^\top\hat\theta=0\). The two terms are \(W\)-orthogonal, \(\hat\theta\) is the minimizer of \eqref{eq:theta-ls}, and \(\dd_0\hat\theta\) is the corresponding fitted flow. If a triangle complex is present, then
\begin{equation}\label{eq:hodge-full}
        r = \delta_1\phi + h,
        \qquad
        h\in \ker(\delta_0)\cap\ker(\dd_1),
\end{equation}
where  \(\delta_1\phi\) is locally cyclic with \(\phi\in C^2\), and \(h\) as  harmonic.
\end{theorem}

\subsection{SigRank statistics}
The normal equations~\eqref{Jiang_LS} yield the centered score
\begin{equation}\label{eq:normal}
        L_W\widehat\theta=B^\top Wy,
        \qquad
        \widehat\theta=L_W^\dagger B^\top Wy.
\end{equation}
Here $^\dagger$ denotes the Moore--Penrose pseudoinverse. The fitted gradient flow is
\[
        B\widehat\theta=P_B^Wy,
        \qquad
        P_B^W=B(B^\top WB)^\dagger B^\top W,
\]
where \(P_B^W\) is the \(W\)-orthogonal projector onto \(\im(B)\).

\begin{definition}[SigRank statistic]\label{def:sigrank}
For \(y\neq0\), define
\begin{equation}\label{eq:T-rank}
        T_G(y)
        =\frac{\norm{P_B^Wy}_W^2}{\norm{y}_W^2}
        =\frac{\norm{B\widehat\theta}_W^2}{\norm{y}_W^2}.
\end{equation}
For \(y=0\), set \(T_G(y)=1\) by convention.
\end{definition}

The statistic \eqref{eq:T-rank} is a descriptive effect size: it is the fraction of observed weighted edge-flow energy explained by the scalar score-difference space. It lies in $[0,1]$, with $T_G=1$ for an exact gradient flow and $T_G=0$ for a flow orthogonal to every score difference, meaning that it provides no information to the flow. The least-squares projection itself is the HodgeRank estimator of \citet{Jiang2011}. SigRank adds graph- and magnitude-conditioned inference for this normalized projection energy. Supplementary Proposition~S4.1 identifies $T_G$ with the uncentered weighted $R^2$ for score regression on the comparison graph; Proposition~S4.2 shows that the exact SigRank $p$-value equals the conditional Gaussian $R^2$ $p$-value given the observed magnitudes.

In our ranking setting, the gradient part is globally rankable, triangle curl measures local cyclic inconsistency, and harmonic flow represents global cycles. Thus $T_G$ defined in \eqref{eq:T-rank} measures the fraction of comparison-flow energy in the gradient component of the Hodge decomposition (Theorem \ref{thm:hodge-projection}). A quite special case where link function is BT, comparison edges are equal weights and fixed $n$, SigRank reduces to the residuals in Equation (13) in \citet{SinghIliopoulosDavidov2025}, but in general, SigRank controls finite-sample false positives under its conditional fair-sign assumption without requiring Gaussian likelihoods.

\begin{corollary} \label{cor:rankability}
Assume \(G\) is connected and \(y\ne 0\). Then:
\begin{enumerate}[label=(\roman*)]
\item \(T_G(y)=1\) if and only if there exists \(\theta\) such that \(y_{ij}=\theta_i-\theta_j\) for every observed edge.
\item \(T_G(y)=0\) if and only if \(B^\top W y=0\), i.e., the edge flow is divergence-free.
\item If triangles are observed in $G$, all triangle sums vanish for the fitted gradient flow:
\[
        (B\hat\theta)_{ij}+(B\hat\theta)_{jk}+(B\hat\theta)_{ki}=0.
\]
\end{enumerate}
\end{corollary}

Thus \(T_G=1\) means exact score-difference representation, whereas \(T_G=0\) means that the flow has no scalar-score projection. Cyclic information remains in the residual \(y-B\widehat\theta\). In practice, this distinction keeps fitted gradient energy separate from residual disagreement and the significance test. If the gradient space fills the observed edge space, every nonzero sign-randomized flow has $T_G=1$, so the conditional test cannot discriminate.

\subsection{Exact conditional randomization validity}

Formally, using the same edge units and conditioning as in the opening statement, the sign null assumes independent Rademacher directions given $G$, $W$, and $a_e=|y_e|$:
\begin{equation}\label{eq:null}
        H_0:\qquad y_e=s_e a_e,
        \qquad s_e\overset{\mathrm{iid}}{\sim}\mathrm{Unif}\{-1,+1\}.
\end{equation}
The alternative is that the observed directions carry more score-difference alignment than $H_0$ would produce:
\begin{equation}\label{eq:alt}
  H_1 : \quad T_G(y) \text{ is larger than expected under } H_0,
\end{equation}
with the test rejecting for large values of $T_G(y)$. This conditional sign null allows heterogeneous observed strengths but assumes independent random directions after conditioning.

To calibrate \eqref{eq:T-rank}, we generate \(B_0\) null replicates $s_e^{(b)}$
$
        y_e^{(b)}=s_e^{(b)}a_e,
        b=1,\ldots,B_0,
$
and compute \(T_G(y^{(b)})\). The Monte Carlo p-value is
\begin{equation}\label{eq:pval}
        p_{B_0}=\frac{1+\#\{b:T_G(y^{(b)})\ge T_G(y)\}}{B_0+1}.
\end{equation}
The 1 in the denominator preserves validity when only a random subset of the sign orbit is sampled \citep{PhipsonSmyth2010,HemerikGoeman2018}. For repeated binary comparisons, the stronger null of independent fair primitive outcomes permits a binomial reference law: condition on \(n_{ij}=N_{ij}+N_{ji}\), draw \(N_{ij}^{(b)}\sim\mathrm{Binomial}(n_{ij},1/2)\) independently across pairs, and transform to margins. This law does not condition on the observed aggregate magnitudes.

\begin{theorem}[Finite-sample validity]\label{thm:exact-validity}
Under \eqref{eq:null}, the exact randomization p-value
\[
        p_{\mathrm{exact}}
        =\Prob\{T_G(s\odot a)\ge T_G(y)\mid a,G,W\},
\]
where $\odot$ denotes entry-wise multiplication, is valid:
\[
        \Prob_{H_0}\{p_{\mathrm{exact}}\le \alpha\mid a,G,W\}\le \alpha
        \quad\text{for every }\alpha\in[0,1].
\]
The Monte Carlo p-value \eqref{eq:pval} is valid in the standard randomized Monte Carlo sense when the observed sign vector is included exchangeably among the simulated sign vectors.
\end{theorem}
For practice, the theorem justifies finite-sample inference on incomplete or unevenly weighted graphs without estimating a response link. The guarantee applies to the declared edge units and conditional-sign law, not merely to marginally fair win rates. In particular, grouping many judgments into one pair margin does not create equally many independent signs, as similar to the properties in SigClust constructions \citep{Liu2008}.

\subsection{Illustrative examples\label{sec:matrix-examples}}
The three examples below illustrate detection of a significant rankable component (Example~\ref{exa:significant-5}), failure to detect when directions are largely cyclic (Example~\ref{exa:insignificant-5}), and the gain from pooling repeated local patterns into a joint test (Section~\ref{sec:multiscale-example}).
\subsubsection{A significant example}
\begin{example}[Five objects with a significant rankable component]\label{exa:significant-5}

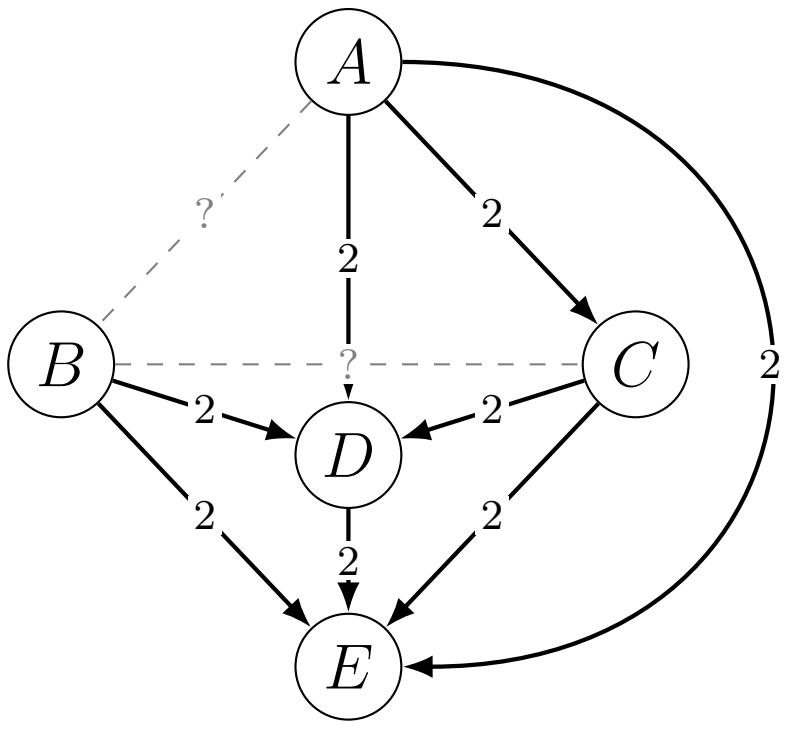
\begin{figure}[h]
\centering
\begin{tikzpicture}[scale=1.0]
  \node[obj] (A) at (0,4) {$A$};
  \node[obj] (B) at (-1.9,2) {$B$};
  \node[obj] (C) at (1.9,2) {$C$};
  \node[obj] (D) at (0,1.4) {$D$};
  \node[obj] (E) at (0,0) {$E$};

  \draw[win] (A) -- node[elab] {$2$} (C);
  \draw[win] (A) -- node[elab] {$2$} (D);
  \draw[win] (A) .. controls (3.6,4) and (3.6,0) .. node[elab,pos=0.5] {$2$} (E);

  \draw[win] (B) -- node[elab] {$2$} (D);
  \draw[win] (B) -- node[elab] {$2$} (E);

  \draw[win] (C) -- node[elab] {$2$} (D);
  \draw[win] (C) -- node[elab] {$2$} (E);

  \draw[win] (D) -- node[elab] {$2$} (E);

  \draw[missing] (A) -- node[elab] {?} (B);
  \draw[missing] (B) -- node[elab] {?} (C);
\end{tikzpicture}
\caption{Significant five-object example.  Solid arrows are observed wins, with edge label \(2\) indicating two repeated wins.  Dashed edges are unobserved comparisons.}
\label{fig:significant-5}
\end{figure}

Let the objects be \(A,B,C,D,E\).  Observe two repeated wins in each of the following eight pair types:
\(A>C\), \(A>D\), \(A>E\), \(B>D\), \(B>E\), \(C>D\), \(C>E\), \(D>E\).
The pairs \(A:B\) and \(B:C\) are unobserved.  Equivalently, the comparison graph is a directed multigraph with sixteen unit-weight directed edges.  Here the sixteen primitive comparison edges, not the eight aggregated pair types, are randomized. Put \(y_e=1\) in the displayed direction for each individual observed edge. Solving \eqref{eq:theta-ls} gives
\[
\begin{array}{c|rrrrr}
\text{object} & A & B & C & D & E\\\midrule
\hat\theta & 0.650 & 0.400 & 0.150 & -0.400 & -0.800
\end{array}
\]
Hence the fitted score order is $A>B>C>D>E$. The explained rankability statistic is $ T_G(y)=0.862500. $ Enumerating all \(2^{16}=65536\) graph-preserving sign flips of the sixteen individual comparison edges yields \(28\) null sign patterns with statistic at least as large.  Thus the conservative exact p-value is $ \frac{28}{65536} = 0.000427246\ldots . $ This is a significant globally orderable component under the sign-flip null.

The conditional central limit theorem (CLT) approximation of Theorem~\ref{thm:TG-clt-null} agrees (standardized statistic \(4.0117\), approximate upper-tail \(p\)-value \(3.01\times10^{-5}\)), although exact enumeration is preferable on this small graph. The ordinary BT maximum likelihood estimator (MLE) is not finite. Supplementary Section~S2.2 gives the calculation.

\end{example}

\subsubsection{An insignificant example}
\begin{example}[The same five objects with no significant rankable component]\label{exa:insignificant-5}

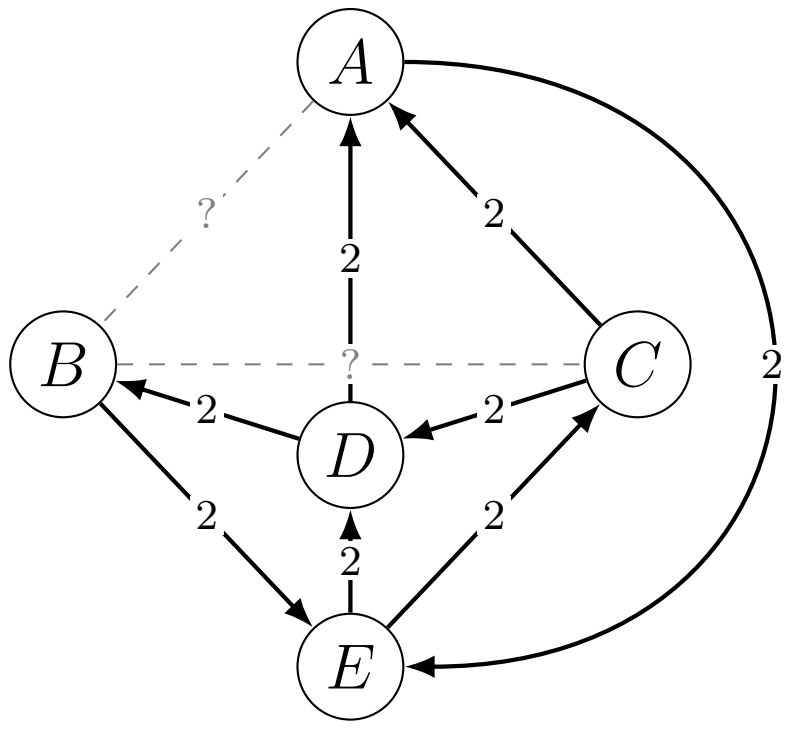
\begin{figure}[h]
\centering
\begin{tikzpicture}[scale=1.0]
  \node[obj] (A) at (0,4) {$A$};
  \node[obj] (B) at (-1.9,2) {$B$};
  \node[obj] (C) at (1.9,2) {$C$};
  \node[obj] (D) at (0,1.4) {$D$};
  \node[obj] (E) at (0,0) {$E$};

  \draw[win] (C) -- node[elab] {$2$} (A);
  \draw[win] (D) -- node[elab] {$2$} (A);
  \draw[win] (A) .. controls (3.6,4) and (3.6,0) .. node[elab,pos=0.5] {$2$} (E);

  \draw[win] (D) -- node[elab] {$2$} (B);
  \draw[win] (B) -- node[elab] {$2$} (E);

  \draw[win] (C) -- node[elab] {$2$} (D);
  \draw[win] (E) to node[elab] {$2$} (C);
  \draw[win] (E) -- node[elab] {$2$} (D);

  \draw[missing] (A) -- node[elab] {?} (B);
  \draw[missing] (B) -- node[elab] {?} (C);
\end{tikzpicture}
\caption{Non-significant five-object example on the same observed comparison graph.  The arrows form several incompatible cycles, so the global score-difference projection explains little of the observed pairwise variation.}
\label{fig:insignificant-5}
\end{figure}

Use the same five objects \(A,B,C,D,E\) and the same observed comparison graph and multiplicities as in the previous example, but change the directions of the eight pair types to
\(C>A\), \(D>A\), \(A>E\), \(D>B\), \(B>E\), \(C>D\), \(E>C\), \(E>D\).
Again, each displayed win is observed twice, so there are sixteen individual unit-weight comparison edges.  The pairs \(A:B\) and \(B:C\) remain unobserved. Solving \eqref{eq:theta-ls} gives
\[
\begin{array}{c|rrrrr}
\text{object} & A & B & C & D & E\\\midrule
\hat\theta & -0.250 & 0.000 & 0.250 & 0.000 & 0.000 .
\end{array}
\]
The least-squares projection returns only a weak score pattern, $C > B=D=E > A$. The explained rankability statistic is only $ T_G(y)=0.062500. $ Enumerating all \(2^{16}=65536\) graph-preserving sign flips gives \(59116\) null sign patterns with statistic at least as large.  Thus the conservative exact p-value is $ \frac{59116}{65536} = 0.902039\ldots . $ Therefore the apparent weak ordering is not significant under the sign-flip null.  Most of the observed variation is cyclic or non-orderable rather than explained by a global score difference.

The conditional CLT agrees (\(Z_{\mathrm{CLT}}=-1.2281\), \(p_{\mathrm{CLT}}=0.8903\)). The BT MLE is finite and unique here. Supplementary Section~S2.3 gives the calculation. MLE existence and evidence for a score-difference component are therefore distinct questions.

\end{example}

\subsubsection{A multiscale example}
\label{sec:multiscale-example}
Let \(G_K\) be a bouquet of \(K\) triangles \(G_\ell =\{O,A_\ell,B_\ell\}\) sharing the anchor \(O\), with unit comparisons \(O>A_\ell>B_\ell\) and \(O>B_\ell\).
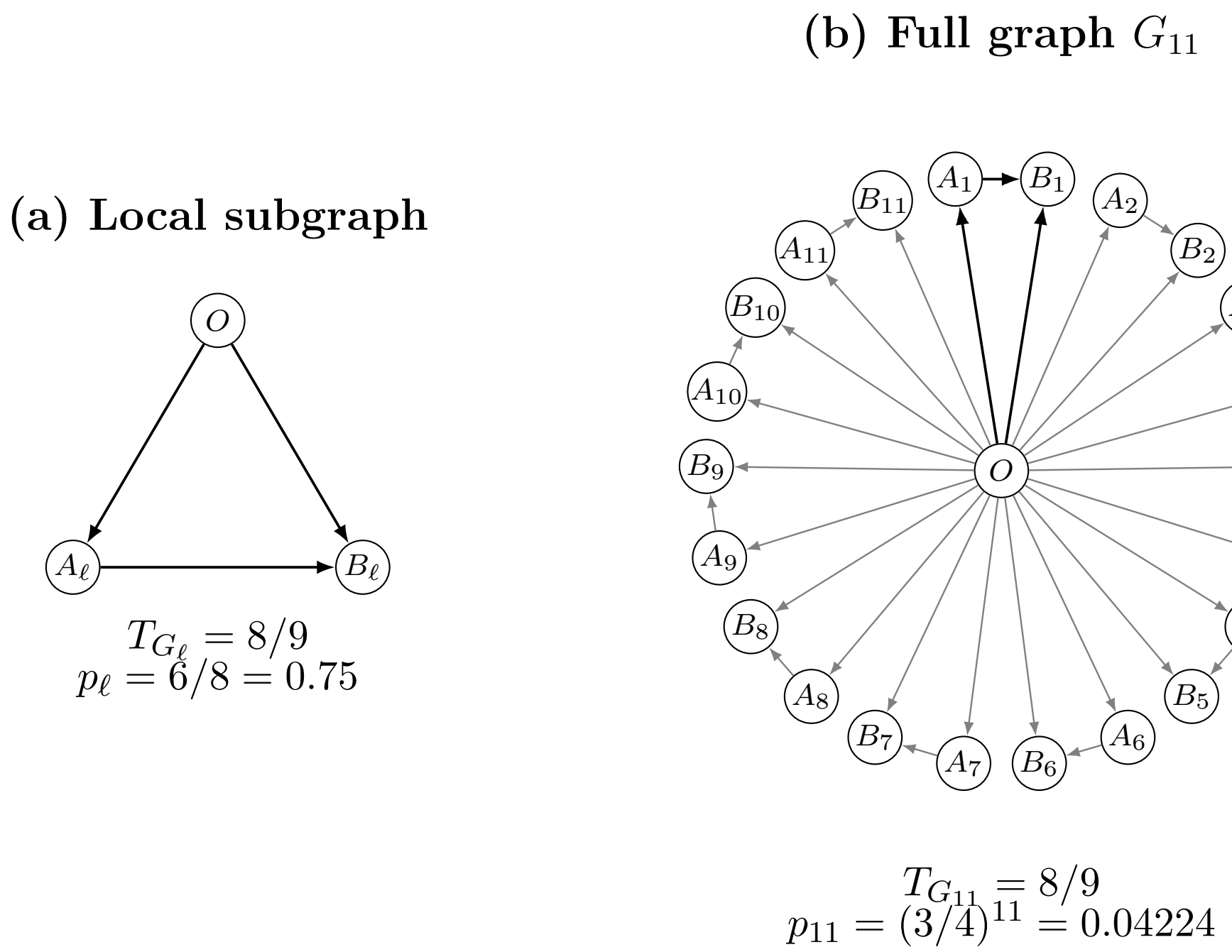
\begin{figure}[h!]
\centering
\begin{tikzpicture}[
    obj/.style={
        circle,
        draw,
        fill=white,
        minimum size=5mm,
        inner sep=0pt,
        font=\scriptsize
    },
    anchor/.style={
        circle,
        draw,
        double,
        fill=white,
        minimum size=7mm,
        inner sep=0pt,
        font=\small
    },
    win/.style={
        -{Latex[length=1.7mm]},
        line width=0.7pt
    },
    repeated/.style={
        -{Latex[length=1.3mm]},
        gray,
        line width=0.45pt
    },
    labeltext/.style={
        align=center,
        font=\small
    }
]

\begin{scope}[xshift=-5.0cm]
    \node[obj] (Olocal) at (0,1.4) {$O$};
    \node[obj] (Alocal) at (-1.35,-0.9) {$A_\ell$};
    \node[obj] (Blocal) at ( 1.35,-0.9) {$B_\ell$};

    \draw[win] (Olocal) -- (Alocal);
    \draw[win] (Alocal) -- (Blocal);
    \draw[win] (Olocal) -- (Blocal);

    \node[labeltext] at (0,-1.75)
    {
        \(T_{G_\ell}=8/9\)\\[-1mm]
        \(p_\ell=6/8=0.75\)
    };

    \node[font=\bfseries] at (0,2.35)
    {(a) Local subgraph};
\end{scope}

\begin{scope}[xshift=2.3cm]
    \node[obj] (O) at (0,0) {$O$};

    \def\R{2.75}

    \foreach \ell in {1,...,11} {
        \pgfmathsetmacro{\midang}{90-(\ell-1)*360/11}
        \pgfmathsetmacro{\aang}{\midang+9.}
        \pgfmathsetmacro{\bang}{\midang-9.}

        \node[obj] (A\ell) at (\aang:\R) {$A_{\ell}$};
        \node[obj] (B\ell) at (\bang:\R) {$B_{\ell}$};

        \draw[repeated] (O) -- (A\ell);
        \draw[repeated] (A\ell) -- (B\ell);
        \draw[repeated] (O) -- (B\ell);
    }

    \draw[win] (O) -- (A1);
    \draw[win] (A1) -- (B1);
    \draw[win] (O) -- (B1);

    \node[font=\bfseries] at (0,4.05)
    {(b) Full graph \(G_{11}\)};

    \node[labeltext,fill=white,inner sep=2pt] at (0,-4.05)
    {
        \(T_{G_{11}}=8/9\)\\[-1mm]
        \(p_{11}=(3/4)^{11}=0.04224\)
    };
\end{scope}

\end{tikzpicture}

\caption{Multiscale rankability in a bouquet of triangles. Each local triangle contains the ranking \(O>A_\ell>B_\ell\), but its exact sign-randomization test is insignificant, with \(p_\ell=0.75\).  When eleven copies of the same local pattern are analyzed jointly on the connected graph \(G_{11}\), the full-graph exact \(p\)-value is \((3/4)^{11}=0.04224\).  The larger-scale test therefore detects a recurring within-subgraph ranking pattern that cannot be established from any individual triangle alone.}
\label{fig:multiscale-bouquet}
\end{figure}
In the cycle orientation, each triangle has flow \(y_\ell=(1,1,-1)^\top\). For a unit-weight triangle \(G\) with nonzero flow \(a=(a_1,a_2,a_3)^\top\),
$
T_G(a)=1-\frac{(a_1+a_2+a_3)^2}
                 {3(a_1^2+a_2^2+a_3^2)},
$
so \(T_G(y_\ell)=8/9\). Six of its eight sign patterns are transitive and attain \(8/9\), giving \(p_\ell=6/8=3/4\).

The \(K\) cycle vectors have disjoint supports, so their residual energies add and \(T_{G_K}=8/9\) as well. Under randomization, the full statistic reaches this value exactly when every triangle is transitive. Hence
\(p_K=(3/4)^K\), \(p_{10}=0.0563\), \(p_{11}=0.04224\).
The common least-squares fit retains \(\widehat\theta_O-\widehat\theta_{A_\ell} =\widehat\theta_{A_\ell}-\widehat\theta_{B_\ell}=2/3\). Thus the joint test detects a recurring family-level pattern at \(K=11\), without making any individual triangle significant.

\section{Statistical properties\label{sec:theory}}
Beyond finite-sample validity, we establish consistency, power, a scalable conditional approximation, and robustness to sign reversals. Proofs are in Supplementary Section~S3.
\subsection{Consistency}
Let a sample-size $n$-dependent graph \(G_n=(V_n,E_n)\) be connected, with \(|V_n|=n\), \(|E_n|=m_n\), incidence matrix \(B_n\), weights \(W_n\), Laplacian \(L_n=B_n^\top W_nB_n\), and projection
\(P_n=B_n(B_n^\top W_nB_n)^\dagger B_n^\top W_n.\)
Assume
\begin{equation}\label{eq:signal-noise}
        y_n=B_n\theta_n^*+\xi_n,
        \qquad \one^\top\theta_n^*=0,
        \qquad \E(\xi_n\mid G_n)=0,
\end{equation}
where the noise coordinates are conditionally independent and satisfy
\[
\E\!\left(\norm{P_n\xi_n}_{W_n}^2\mid G_n\right) \le C_\xi(n-1).
\]

\begin{theorem}[Observed-edge Hodge consistency]\label{thm:consistency}
Let \(\widehat\theta_n\) solve \eqref{eq:theta-ls}. Under \eqref{eq:signal-noise},
\begin{equation}\label{eq:observed-edge-consistency}
        \norm{B_n(\widehat\theta_n-\theta_n^*)}_{W_n}^2=O_p(n).
\end{equation}
Consequently,
\(\frac1{m_n} \norm{B_n(\widehat\theta_n-\theta_n^*)}_{W_n}^2 =O_p\!\left(\frac{n}{m_n}\right),\)
which converges to zero when \(m_n/n\to\infty\). Moreover, if \(\lambda_2(L_n)\) denotes the smallest positive eigenvalue of the connected weighted Laplacian, then
\begin{equation}\label{eq:score-error-consistency}
        \norm{\widehat\theta_n-\theta_n^*}_2^2
        \le
        \frac{
        \norm{B_n(\widehat\theta_n-\theta_n^*)}_{W_n}^2
        }{\lambda_2(L_n)}
        =O_p\!\left(\frac{n}{\lambda_2(L_n)}\right).
\end{equation}
Consequently, the average squared score error is \(O_p\{1/\lambda_2(L_n)\}\) and converges to zero when \(\lambda_2(L_n)\to\infty\).
\end{theorem}

\begin{theorem}[Power consistency of the randomization test]\label{thm:power}
Suppose \eqref{eq:signal-noise} holds and \(m_n/n\to\infty\). Assume uniformly bounded edge weights and uniformly bounded second moments for the coordinates of \(W_n^{1/2}\xi_n\), together with
\[
        c m_n\le
        S_n:=\norm{B_n\theta_n^*}_{W_n}^2
        \le C m_n
\]
for constants \(0<c\le C<\infty\). Assume also $ \max_{e\in E_n}w_{n,e}y_{n,e}^2=O_p(1). $ Then \(T_{G_n}(y_n)\) is bounded away from zero in probability, whereas a conditional sign-flip replicate satisfies
\[
        T_{G_n}(y_n^{(b)})=O_p(n/m_n).
\]
Hence the exact randomization test has power tending to one.
\end{theorem}
Estimation and testing require different assumptions. The signal-plus-noise model in \eqref{eq:signal-noise} describes alternatives for these consistency results. Finite-sample null validity does not require it. Theorem~\ref{thm:consistency} controls prediction error on observed edges. Converting this to score accuracy depends on the weighted Laplacian gap, so weakly connected designs can estimate observed comparisons more accurately than individual scores. Ordering recovery and winner stability require additional control of score gaps and perturbations. The power theorem gives a sufficient regime in which the edge-to-vertex ratio grows and signal energy remains substantial under the stated bounds. It does not imply high power for sparse graphs or for many judgments aggregated into few edge units.

\subsection{Asymptotics}\label{sec:asymptotics}

The exact sign-flip test is finite-sample valid.  For large graphs, one can also derive a Gaussian approximation.  Write the observed edge flow as
\(y=s\odot a\), \(s_e\in\{-1,+1\}\), \(a_e=|y_e|\).
Write $P:=P_B^W$, let $b_e=\sqrt{w_e}\,a_e$ and $D_b=\operatorname{diag}(b_e)$, and let $\Pi=W^{1/2}PW^{-1/2}$ be the Euclidean orthogonal projection corresponding to the \(W\)-orthogonal projection \(P\) onto \(\operatorname{im}(B)\).  Then
\[
        \|Py\|_W^2
        =
        s^\top Hs,
        \qquad
        H=D_b\Pi D_b,
 \qquad
        \|y\|_W^2
        =
        \sum_{e\in E} b_e^2
        =:A.
\]
Therefore, under the sign-flip null \eqref{eq:null},
$
        T_G(y)=\frac{s^\top Hs}{A}.
$

\begin{theorem}[Null CLT]
\label{thm:TG-clt-null}
Consider a sequence of conditioning values \((a_n,G_n,W_n)\), and condition on each triple. Let $s_n=(s_{n,e})_{e\in E_n}$ have independent Rademacher entries under the sign-flip null, and define
\[
 Q_n=s_n^\top H_ns_n,\qquad \mu_n=\operatorname{tr}(H_n),\qquad K_n=H_n-\operatorname{diag}(H_n).
\]
Then $Q_n-\mu_n=s_n^\top K_ns_n$. Let
\[
        \sigma_n^2
        =
        \operatorname{Var}(Q_n\mid a_n,G_n,W_n)
        =
        4\sum_{e<f}H_{n,ef}^2
        =
        2\operatorname{tr}(K_n^2).
\]
Assume \(\sigma_n>0\) for all sufficiently large \(n\) and suppose the following two negligibility conditions hold:
\[
        \frac{
        \max_{e\in E_n}\sum_{f\ne e}K_{n,ef}^2
        }{
        \sigma_n^2
        }
        \longrightarrow 0,
        \qquad
        \frac{\|K_n\|_{\mathrm{op}}^2}{\sigma_n^2}
        \longrightarrow 0.
\]
Here $\|\cdot\|_{\mathrm{op}}$ denotes the Euclidean operator norm. Then, along any sequence of conditioning values satisfying these assumptions,
\[
        \frac{Q_n-\mu_n}{\sigma_n}
        \Rightarrow N(0,1)
\]
under the conditional sign-flip distribution.

Equivalently, since
\(Q_n=A_nT_{G_n}(y_n)\), \(A_n=\sum_{e\in E_n}w_{n,e}a_{n,e}^2\),
we have
\[
        \frac{
        A_nT_{G_n}(y_n)-\operatorname{tr}(H_n)
        }{
        \sigma_n
        }
        \Rightarrow N(0,1).
\]
Thus, with $\Phi$ denoting the standard normal distribution function, the corresponding asymptotic upper-tail approximation is
\[
        p_{\mathrm{CLT}}
        =
        1-\Phi\left(
        \frac{
        A_nT_{G_n}(y_n)-\operatorname{tr}(H_n)
        }{
        \sigma_n
        }
        \right).
\]
\end{theorem}

The central limit theorem approximates the conditional null law for the particular graph, weights, and magnitudes. Its negligibility assumptions exclude concentration of randomization variance in one edge's interactions or one eigen-direction of the off-diagonal kernel. A large edge count alone is therefore insufficient. Centering at $\tr(H_n)$ is essential: a large projection fraction needs not be unusual for its own conditional null. Multiplying all margins by a common nonzero factor changes neither the standardized statistic nor these relative conditions. If the conditional variance vanishes, the statistic is constant over its sign orbit and the standardized approximation is undefined. When variance is positive but the asymptotic conditions are doubtful, exact enumeration or Monte Carlo calibration avoids the Gaussian approximation. %

Supplementary Proposition~S3.2 characterizes first-order agreement between conditional SigRank and unconditional Gaussian $R^2$ critical values \citep{SinghIliopoulosDavidov2025}. In balanced designs with equal weighted edge energies and uniform projection leverages, the null means coincide and the variance ratio tends to one. This asymptotic comparison is distinct from the exact conditional Gaussian correspondence in Supplementary Proposition~S4.2.
Supplementary Section~S4 gives the regression and BT score embeddings.

\subsection{Robustness}\label{sec:robustness}

The unpenalized BT MLE is finite and unique up to an additive constant only under the classical strong-connectivity condition on the directed win graph \citep{Ford1957}. For this BT comparison, let $t=1,\ldots,N$ index primitive comparisons between items $i_t$ and $j_t$, let $x_t\in\{0,1\}$ indicate a win of $i_t$, and set $b_t=e_{i_t}-e_{j_t}$, where $e_i$ is the $i$th coordinate vector. The data are completely separated if there exists \(v\in\one^\perp\) such that $(2x_t-1)b_t^\top v>0$ for every $t$. The first example in Section~\ref{sec:matrix-examples} is separated: its BT MLE is not finite, whereas SigRank remains well defined and exactly calibrated.

\subsubsection{Average-case sign errors}

Fix \(G,W\) and a nonzero observed flow \(y\), so \(A>0\). Expectations and variances are over the corruption variables. We retain the notation of Section~\ref{sec:asymptotics}:
\(z=W^{1/2}y=D_bs\), \(H=D_b\Pi D_b\), \(Q:=Q(z):=z^\top\Pi z=s^\top Hs\), \(A:=\norm{z}_2^2\), \(T_G(y)=Q/A\).
Under the conditional null,
$\mu_H:=\E_0 Q=\tr(H),
        \qquad
        \sigma_H^2:=\Var_0(Q)=4\sum_{e<f}H_{ef}^2$.
When \(\sigma_H>0\), write
\begin{equation}\label{eq:robust-Z}
        Z=\frac{Q-\mu_H}{\sigma_H}.
\end{equation}
These null quantities depend only on the graph, weights, and magnitudes and therefore remain fixed under sign corruption.

Let \(\xi_1,\ldots,\xi_m\) be independent with
\(\Prob(\xi_e=-1)=\varepsilon\), \(\Prob(\xi_e=1)=1-\varepsilon\), \(0\le\varepsilon\le\tfrac12\),
and define \(s^{(\varepsilon)}=s\odot\xi\) and \(r=1-2\varepsilon\).

\begin{theorem}[Independent sign-error attenuation and variance]\label{thm:robust-independent}
Let \(Q_\varepsilon=(s^{(\varepsilon)})^\top Hs^{(\varepsilon)}\), \(T_\varepsilon=Q_\varepsilon/A\), and \(Z_\varepsilon=(Q_\varepsilon-\mu_H)/\sigma_H\). Then
\begin{align}
        \E Q_\varepsilon
        &=\mu_H+r^2(Q-\mu_H),
        \label{eq:robust-meanQ}\\
        \E T_\varepsilon
        &=\frac{\mu_H}{A}
          +r^2\left(T_G-\frac{\mu_H}{A}\right),
        \label{eq:robust-meanT}\\
        \E Z_\varepsilon
        &=r^2Z.
        \label{eq:robust-meanZ}
\end{align}
Define \(C_{ef}=s_es_fH_{ef}\) for \(e\ne f\) and \(C_{ee}=0\). Then
\begin{align}
\Var(Q_\varepsilon)
=4\bigg[
  (1-r^2)^2\sum_{e<f}C_{ef}^2  +r^2(1-r^2)
 \sum_e\left(\sum_{f\ne e}C_{ef}\right)^2
\bigg].
\label{eq:robust-var}
\end{align}
Consequently, \(\Var(T_\varepsilon)=\Var(Q_\varepsilon)/A^2\).
\end{theorem}

The expected excess standardized signal therefore contracts exactly by \((1-2\varepsilon)^2\), not merely asymptotically. This quantifies sensitivity to a specified independent label-error rate, but it does not by itself give the corrupted test's power or Type-I error.

\subsubsection{Worst-case sign errors}

For a deterministic set \(S\subseteq E\), let \(F_S\) be the diagonal matrix with entry one on \(S\) and zero elsewhere, and define
$
        z^{(S)}=(I-2F_S)z,
        T^{(S)}=Q(z^{(S)})/A.
$
Let
\begin{equation}\label{eq:robust-rho-eta}
        \rho(S)=\frac{\norm{F_Sz}_2^2}{A},
        \qquad
        \eta(S)=\frac{\norm{\Pi F_Sz}_2^2}{A}
        \le\rho(S).
\end{equation}

For a vector $u$, let $u_S$ denote its restriction to $S$. Let $\Pi_{SS}$ denote the corresponding principal submatrix, and write $x_+=\max\{x,0\}$.

\begin{theorem}[Exact perturbation identity and deterministic bounds]\label{thm:robust-adversarial}
For every \(S\subseteq E\),
\begin{equation}\label{eq:robust-identity}
        Q(z^{(S)})-Q(z)
        =-4z_S^\top(\Pi z)_S
         +4z_S^\top\Pi_{SS}z_S.
\end{equation}
Moreover,
\begin{equation}\label{eq:robust-eta-bound}
\left(\sqrt{T_G}-2\sqrt{\eta(S)}\right)_+^2
\le T^{(S)}
\le
\min\left\{1,
\left(\sqrt{T_G}+2\sqrt{\eta(S)}\right)^2
\right\},
\end{equation}
and in particular
\begin{equation}\label{eq:robust-rho-bound}
        T^{(S)}
        \ge
        \left(\sqrt{T_G}-2\sqrt{\rho(S)}\right)_+^2.
\end{equation}
\end{theorem}

\begin{corollary}[Single-edge influence]\label{cor:robust-certificate}
If only edge \(e\) is reversed, write $T_G^{(e)}:=T^{(\{e\})}$. Then
\begin{equation}\label{eq:robust-single}
        T_G^{(e)}-T_G
        =\frac{-4z_e(\Pi z)_e+4z_e^2\Pi_{ee}}{A}.
\end{equation}
Let \(c_\alpha\) be an upper-tail conditional critical value, with rejection when \(T_G>c_\alpha\). If \(T_G>c_\alpha\), every set \(S\) satisfying
\begin{equation}\label{eq:robust-certificate}
        \rho(S)
        <\frac{(\sqrt{T_G}-\sqrt{c_\alpha})^2}{4}
\end{equation}
is guaranteed to preserve rejection. For a budget of at most \(q\) errors, where \(0\le q\le m\), define
\(\rho_q^*=\max_{|S|\le q}\rho(S).\)
Condition \eqref{eq:robust-certificate} with \(\rho_q^*\) certifies rejection against every set of at most \(q\) errors. If \(w_ey_e^2\) is constant over edges, then \(\rho_q^*=q/m\). This provides an interpretable worst-case guarantee in units of weighted edge energy, rather than an unconditional guarantee against deleting individual judgments or changing a fitted winner.
\end{corollary}

The robustness statements hold the graph, weights, and magnitudes fixed under sign reversals, preserving the conditional reference law and its critical value. Average-case attenuation describes the excess above the conditional null mean, rather than a fixed proportional decrease in the entire statistic or its power. The deterministic certificate protects rejection against every allowed reversal set, whose cost is its share of weighted flow energy. A few high-energy edges may therefore matter more than many low-energy edges. These guarantees do not directly cover deleting individual judgments, which can change weights and margins, or changes in the fitted winner. The BT bounds additionally require finite fits and positive likelihood curvature.

For comparison, let \(X\) have rows \(b_t^\top\) and define the centered BT negative log-likelihood
\begin{equation}\label{eq:BTloss}
        \ell_x(\theta)
        =\sum_{t=1}^N
        \left[
        \log\{1+\exp(b_t^\top\theta)\}
        -x_tb_t^\top\theta
        \right].
\end{equation}
When one binary label is used per observed matrix edge, reversing \(h\) labels is the same edge-sign error considered above. Supplementary Theorem~S3.5 and Corollary~S3.6 give finite-fit perturbation bounds, and Proposition~S3.7 shows that symmetric outcome errors need not preserve the BT family.

\section{Simulation Studies}
\label{sec:simulation}

\subsection{Hypotheses and design}
\label{sec:simulation-design}

We compare the BT equal-score test with SigRank's conditional test from Section~\ref{sec:matrix-sigrank}, at nominal level $\alpha=0.05$. Let $N_e$ count wins in the reference direction among $r$ primitive comparisons on pair $e$, and let $\vartheta$ denote the BT score vector. Within the independent-binomial BT model, $N_e\sim\operatorname{Binomial}\{r,\sigma((B\vartheta)_e)\}$, where $\sigma(t)=(1+e^{-t})^{-1}$, the score test compares
\[
 H_0^{\mathrm{BT}}:B\vartheta=0
 \qquad\text{against}\qquad
 H_1^{\mathrm{BT}}:B\vartheta\ne0.
\]
SigRank instead tests the conditional sign null in~\eqref{eq:null},
\[
 H_0^{\mathrm{SR}}:
 y=s\odot|y|,\qquad s_e\overset{\mathrm{iid}}{\sim}
 \operatorname{Unif}\{-1,+1\}\quad\text{given }(G,W,|y|),
\]
against unusually strong score-difference alignment, as in~\eqref{eq:alt}. It does not test goodness of fit of the BT model. We fix 18 items, 54 observed pairs, and $r=9$ comparisons per pair. The marginal win probability is $\mu_e=\sigma\{\beta(\theta_i-\theta_j)\}$ for a fixed standardized score vector $\theta$, with $\beta\ge0$ controlling signal strength. Independent beta-binomial counts across pairs introduce within-pair correlation $\tau_e$. Setting $\tau_e=0$ recovers binomial sampling. Both tests use $y_e=(2N_e-r)/r$, $W=rI$, and projection energy $Q_G=\|P_B^Wy\|_W^2$. BT simulates its independent fair-binomial reference law directly. SigRank randomizes the aggregated pair signs conditional on their magnitudes. Thus the BT comparison does not rely on a chi-square approximation.

The independent-binomial BT null implies SigRank’s conditional fair-sign null. Conditioning the BT score statistic on the observed margin magnitudes yields the same exact sign-randomization tail as SigRank. At \(\tau_e=0\), both assumptions hold, providing a common baseline for calibration and power. Positive within-pair correlation preserves the conditional sign null while changing the magnitude distribution assumed by the BT reference. These settings therefore assess robustness of the nominal BT calibration to the specified sampling misspecification. %

Throughout, Type-I error means rejection under the generating null $\beta=0$, and power means rejection under $\beta>0$, using $p\le0.05$. At $\beta=0$, independent symmetric pair counts satisfy SigRank's conditional null for every $\tau_e$, whereas positive $\tau_e$ violates BT's independent-binomial sampling law. Supplementary Section~S5.1 gives the generator and calibration details.

We use $5{,}000$ datasets per null setting and $2{,}000$ per positive signal $\beta\in\{0.12,\allowbreak0.24,\allowbreak0.36,\allowbreak0.48,\allowbreak0.60\}$. Type-I error and power are the proportions with $p\leq0.05$. Marginal 95\% Wilson intervals describe Monte Carlo uncertainty conditional on the fixed graph. Both procedures are evaluated on the same datasets, against these same null and alternative settings. Their different assumptions determine how they calibrate evidence for that signal.

\subsection{Experiment 1: increasing within-pair correlation}
\label{sec:simulation-correlation}

\begin{figure}[t!]
 \centering
 \includegraphics[width=\linewidth]{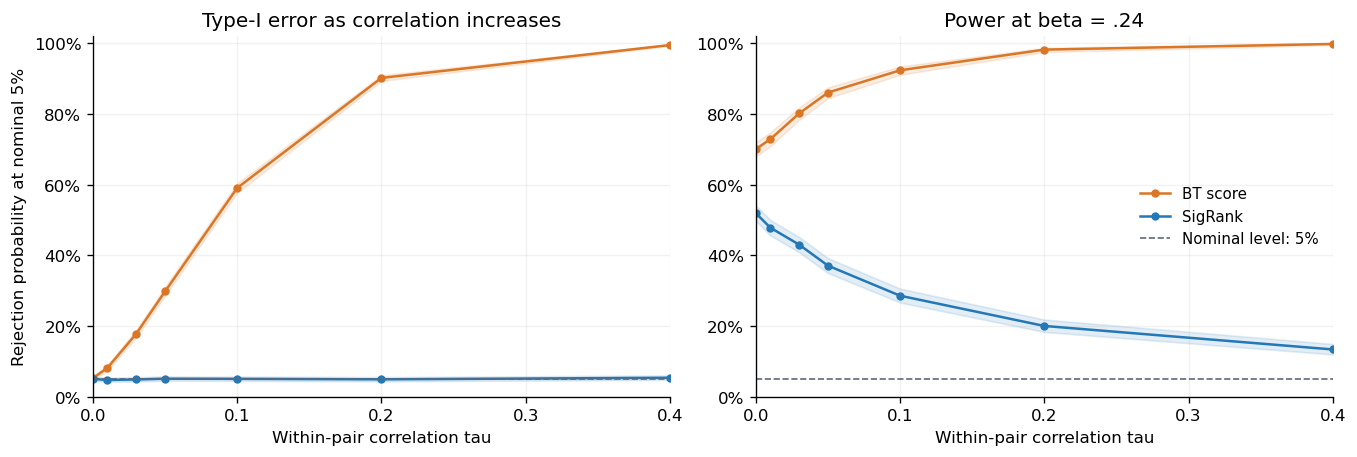}\par
 \smallskip
 {\small
 \begin{tabular}{rcccc}
 \toprule
 & \multicolumn{2}{c}{Type-I error (\%)} & \multicolumn{2}{c}{Power (\%)} \\
 \cmidrule(lr){2-3}\cmidrule(lr){4-5}
 $\tau$ & BT & SigRank & BT & SigRank \\
 \midrule
 $0$ & 5.22 & 5.02 & 70.00 & 51.90 \\
 $0.10$ & 59.02 & 5.04 & 92.30 & 28.55 \\
 $0.40$ & 99.44 & 5.34 & 99.75 & 13.35 \\
 \bottomrule
 \end{tabular}
 }
 \caption{Experiment 1: empirical Type-I error and power at $\beta=0.24$ as within-pair correlation $\tau$ increases, using $p\le0.05$. Top: Type-I error and power curves, with the dashed reference at nominal 5\%. Bottom: selected numerical operating points in percent. Shading gives marginal 95\% Wilson Monte Carlo intervals.}
 \label{fig:simulation-calibration}
\end{figure}

We set $\tau_e=\tau\in\{0,0.01,0.03,0.05,0.10,0.20,0.40\}$. Figure~\ref{fig:simulation-calibration} shows the correlation sweep and selected operating points at $\beta=0.24$. Supplementary Figure~S1 shows receiver operating characteristic curves at $\tau=0.10$, and Supplementary Table~S1 gives the areas under the curves for each positive $\beta$ at $\tau=0$ and $0.10$. At $\tau=0$, Type-I errors are 5.22\% (BT) and 5.02\% (SigRank). BT's null rejection rate reaches 59.02\% at $\tau=0.10$ and 99.44\% at $\tau=0.40$. SigRank remains between 4.72\% and 5.34\%. Within-pair dependence multiplies the null count variance by $1+8\tau$, which the BT reference omits. SigRank preserves the observed magnitudes and calibrates their directional alignment.

Under the correctly specified model, BT has higher power in this design. At $\beta=0.24$, its baseline power is 70.00\%, versus 51.90\% for SigRank. SigRank power falls to 13.35\% at $\tau=0.40$: preserving calibration does not prevent information loss. BT's larger alternative rejection rates under misspecification must be read with its inflated null rates.

\subsection{Experiment 2: common and heterogeneous dispersion}
\label{sec:simulation-cases}

\begin{figure}[t!]
 \centering
 \includegraphics[width=0.95\linewidth]{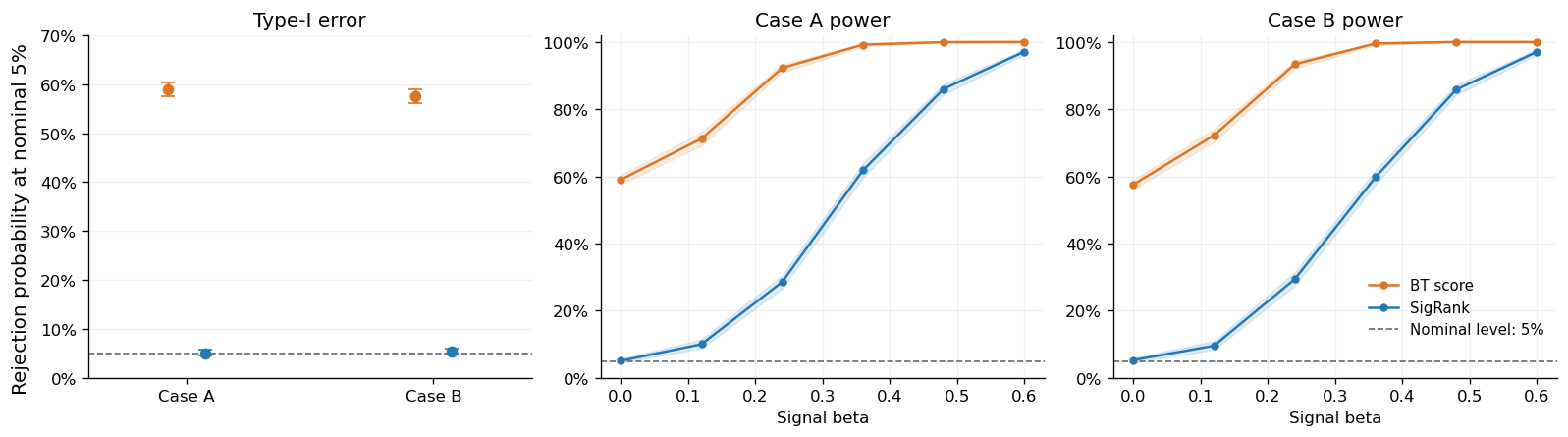}
\caption{Experiment 2: null rejection rates and power in Cases A and B at nominal level 5\%. Power-curve endpoints at $\beta=0$ are null rejection rates. Bars and shading show marginal 95\% Wilson Monte Carlo intervals. Case A reuses Experiment 1's $\tau=0.10$ datasets.}
 \label{fig:simulation-cases}
\end{figure}

Case A sets $\tau_e=0.10$ throughout. Case B fixes $\tau_e\in\{0,0.05,0.25\}$ on 18 pairs each. Both have average null count variance 4.05, but Case B has unequal pair variances. Figure~\ref{fig:simulation-cases} shows BT null rejection rates of 59.02\% and 57.56\%, versus SigRank's 5.04\% and 5.26\%. At $\beta=0.24$, SigRank power is 28.55\% and 29.40\%, respectively, and increases with signal. Case B combines heterogeneous dispersion with within-pair dependence. The small A/B difference does not isolate an effect of heterogeneity. SigRank's calibration here follows from the preserved conditional sign null, not from likelihood-free construction alone.

\section{Applications}
\label{sec:llm-arena}

The Chatbot Arena tensor contains 22,998 comparisons among 20 models in 20 language slices \citep{Chiang2024ChatbotArena}. The retrospective primary analysis covers all languages and four common-order groups, with Holm correction across 24 tests (Holm-24) \citep{Holm1979}: 13 detections, 10 %
non-rejections, and one comparison limited by exact resolution. We focus on what agreement and disagreement between BT and SigRank reveal about the observed data. Supplementary Sections~S6 and~S5.3 give the design and additional diagnostics.

\subsection{Hypotheses and design}
\label{subsec:arena-data}
\label{subsec:arena-family-sigrank}

Within each group, one language--pair margin is one SigRank edge unit, weighted by its comparison count. Language-specific margins remain parallel edges, and the projection fits one shared score vector. BT tests equal shared scores under independent binomial trials. SigRank tests the conditional fair-sign null~\eqref{eq:null}, holding the graph, counts, and absolute margins fixed. Thus 164 judgments need not provide 164 independent sign units.

The matched secondary analysis applies Holm separately to each method across the same four count-weighted groups (Holm-4). Both reject for LLaMA, only BT rejects for GPT and Pythia, and neither rejects for Claude, whose five nonzero signs cannot resolve nominal 5\%. Equal-language SigRank sensitivity and exact binomial checks use separate correction families, detailed in the appendix.

For a model group $F$, write $G_F^\star$ for its common-order graph with $m_F$ language--pair edges. Let $B_F^\star$, $W_F^\star$, and $y_F^\star$ denote its incidence matrix, diagonal edge-weight matrix, and margin vector. Set $P_F^\star:=P_{B_F^\star}^{W_F^\star}$ and $T_F^\star:=T_{G_F^\star}(y_F^\star)$, using the definitions in Section~\ref{sec:matrix-sigrank}.

The conditional null mean describes each group's geometry. Let $\Pi_F^\star=(W_F^\star)^{1/2}P_F^\star(W_F^\star)^{-1/2}$, $h_e=(\Pi_F^\star)_{ee}$, and $v_e=w_ey_e^2$. Independent fair signs cancel the cross terms, giving
\[
 \E_s(T_F^\star\mid G_F^\star,W_F^\star,|y_F^\star|)
 =\frac{\sum_e h_ev_e}{\sum_e v_e}.
\]
Here $h_e$ is the edge leverage in the whitened score regression. The leverages lie between zero and one and sum to $\rank(B_F^\star)$. Equal positive edge energies therefore recover the dimension ratio $\rank(B_F^\star)/m_F$, whereas heterogeneous energies weight the leverages differently. Energy concentrated on high-leverage coordinates can produce a large conditional null mean. The reported means summarize this interaction between geometry and observed energy.

Count weights emphasize languages with more comparisons. The separate equal-language sensitivity analysis uses $w_{\ell ij}^{\mathrm{bal}}=n_{\ell ij}/\sum_{(a,b)\in E_{\ell F}}n_{\ell ab}$, where $\ell$ indexes language, $n_{\ell ij}$ is its comparison count for pair $(i,j)$, and $E_{\ell F}$ contains that language's observed group pairs. Each represented language then has total weight one, retaining relative counts within it. Reweighting changes the target and conditional reference law, requiring recalibration. These balanced weights are used only for SigRank sensitivity, while BT uses original counts as binomial trials.

The matched comparison reuses the existing SigRank tails. BT tests equal scores with $Q_G$ and its asymptotic $\chi^2_{\rank(B)}$ law under independent binomial trials. The matched Holm-4 values are distinct from primary Holm-24 values. Exact binomial checks for all three two-model groups have their own correction across three tests (Holm-3). No joint error-control claim spans these families or methods. All BT fits are finite, which does not establish adequacy. The comparison is retrospective and provides no independent confirmation.

\subsection{LLaMA: corroborating evidence}
The eight-model LLaMA ecosystem has 2,684 comparisons and 169 language-pair edges, including 162 nonzero sign units. Its common score explains 79.8\% of observed energy, compared with a conditional null mean of 26.4\%. None of 199,999 sign draws reaches the observation. The add-one value at the Monte Carlo floor, $p=0.000005$, gives primary Holm-24 $p=0.000120$ and matched Holm-4 $p=0.000020$. BT also rejects equal scores ($Q_G=528.5422$, Holm-4 $p=2.33\times10^{-109}$). Equal-language weighting retains the SigRank detection ($T_F^\star=0.4897$, Holm-4 $p=0.000020$). This is evidence for a detectable component across the observed language-pair units, without establishing a universal order or exact additivity.

\begin{figure}[!t]
\centering
\includegraphics[width=\textwidth]{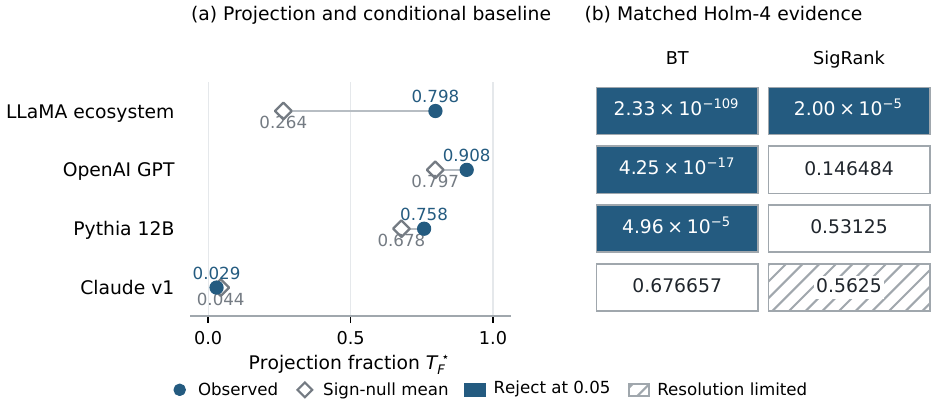}
\caption{Count-weighted common-order results. (a) Observed projection fractions and conditional sign-null means. These means are not critical values. (b) Matched Holm-4 p-values. Filled cells reject at 5\%. Hatched Claude has minimum exact SigRank $p=0.0625$. LLaMA's adjusted SigRank value derives from the raw add-one floor $5\times10^{-6}$ (199,999 Monte Carlo draws). The model groups and weighting schemes are specified in Supplementary Section~S6.2.}
\label{fig:arena-evidence-summary}
\end{figure}

\subsection{Pythia: a BT contrast without SigRank corroboration}
\label{subsec:pythia-sigrank}
Pythia compares \texttt{oasst-pythia-12b} with \texttt{dolly-v2-12b}: 164 comparisons across nine languages, with 109 wins for the first model and fitted BT probability $0.665$. BT rejects equal scores with $Q_G=17.7805$, raw $p=2.48\times10^{-5}$, and Holm-4 $p=4.96\times10^{-5}$. The exact binomial check also rejects (raw $p=2.99\times10^{-5}$, three-check Holm $p=5.99\times10^{-5}$), so this conclusion does not depend on the chi-square approximation.

SigRank has seven nonzero sign units and 64 configurations after global-sign compression, which identifies each sign pattern with its complete reversal because both have the same statistic. Its projection fraction is 75.8\%, but its conditional null mean is already 67.8\%. The exact tail $17/64=0.265625$ becomes $0.53125$ under matched Holm-4 adjustment and $1$ under primary Holm-24 adjustment. Equal-language weighting also does not reject ($T_F^\star=0.0703$, adjusted $p=0.53125$). Thus pooled count evidence is not accompanied by detected conditional sign coherence. The minimum exact $p=1/64$ permits nominal rejection but establishes neither adequate power after correction nor absence of ranking structure.

This contrast has a concrete data interpretation. The pooled sample favors the first model, but the observed language margins include opposing directions. Under equal-language weights the projection fraction falls to 0.0703. If all nonzero margins shared a sign, the statistic would be at least $1/9$ (Supplementary Section~S5.3.3). Thus the margins partly cancel when languages receive equal weight. The high count-weighted null mean also reflects concentration on high-leverage, high-count language units. A large pooled projection can consequently arise without unusual alignment of the observed signs.

These comparisons concern different reference laws. BT accumulates evidence from comparison counts. SigRank asks whether directions align unusually well after conditioning on their magnitudes. Missing SigRank corroboration warrants caution before extending a fitted BT contrast to a broad common-ranking claim. SigRank non-rejection is not a BT goodness-of-fit rejection. Language support, dependence, model fit, and robustness require separate assessment. Prompt, user, and time identifiers are unavailable here.

Even agreement does not ensure a stable winner. In English, SigRank detects structure, yet deleting 139 of 20,670 effective judgments (0.672\%) changes the refitted BT winner from \texttt{gpt-4} to \texttt{claude-instant-v1}. The pooled  result changes the same winner after 164 of 22,998 (0.713\%). These are verified witnesses along an approximate maximum influence perturbation (AMIP) path within the declared one-percent budgets, not global deletion minima or named source records.

\section{Discussion}

SigRank tests whether the chosen observed edge flow has an unusually large score-difference component under the conditional sign null. The statistic $T_G$ is a descriptive weighted graph $R^2$, while its $p$-value depends on the declared edge units, graph, weights, magnitudes, and joint fair-sign assumption. In applications, report both the effect size and the calibrated evidence, and distinguish either from agreement with a fitted parametric ranking model.

BT and Thurstone share this latent score-difference geometry but impose response links that need not preserve additivity of the observed margin flow. The same distinction applies to BTL models \citep{BradleyTerry1952,luce1959individual}. The BT score embedding in Supplementary Proposition~S4.3 makes the distinction explicit: the same projection energy can be calibrated under different reference laws. When model specification is uncertain or within-pair dependence cannot be ruled out, SigRank provides a valid first-stage check for score-difference structure before committing to a parametric ranking model.

Validity of the base test requires the conditional sign null for the chosen edge representation. Primitive-record and aggregated-pair analyses answer different conditional questions. Weights and support are treated as fixed in the base theory, and the sign-reversal sensitivity bounds depend on the perturbation law being scientifically meaningful. The retrospective Chatbot Arena analysis accounts for its declared family, but broader adaptive searches require corresponding selection-adjusted calibration.
\vspace{1cm}

\noindent\textbf{Acknowledgement.} The authors thank Yiming Xu for helpful discussions and comments for this research manuscript.

\noindent\textbf{Conflict of Interest.} There is no conflict of interest.

\noindent\textbf{Supplementary materials.} The supplementary materials contain proofs of the theoretical results, additional graph-theoretic calculations and illustrative examples, comparisons with Gaussian and Bradley--Terry calibration, and details of the Chatbot Arena data representation, model groups, and weighting schemes.

\medskip
\noindent\textbf{Data availability statement.} The aggregated Chatbot Arena comparison tensor analyzed in this study is included in the accompanying reproducibility package, together with the associated analysis code and computational instructions. The tensor contains language-specific pairwise comparison counts and margins derived from Chatbot Arena \citep{Chiang2024ChatbotArena}.

\medskip
\noindent\textbf{Declaration of Generative AI Use.} The authors used OpenAI ChatGPT-6 Sol to assist with manuscript drafting and revision and code development (including algorithm design and figure preparation) and debugging. AI assistance in the mathematical material included proofreading of Theorem~\ref{thm:robust-independent}, %
 and drafting the Gaussian-calibration derivation in Supplementary Proposition~S4.2. The authors reviewed and revised the AI-assisted material, verified the mathematical statements, references, code, and numerical results for suitability for publication. The authors take full responsibility for the originality, accuracy, and integrity of the article.

\bibliographystyle{plainnat}
\bibliography{sigrank_refs}

@article{AndersonChartierLangville2019,
  author = {Paul E. Anderson and Timothy P. Chartier and Amy N. Langville},
  title = {The rankability of data},
  journal = {SIAM Journal on Mathematics of Data Science},
  volume = {1}, number = {1}, pages = {121--143}, year = {2019},
  doi = {10.1137/18M1183595}
}

@article{AndersonEtAl2021Weighted,
  author = {Paul E. Anderson and Timothy P. Chartier and Amy N. Langville and Kathryn E. Pedings-Behling},
  title = {The rankability of weighted data from pairwise comparisons},
  journal = {Foundations of Data Science},
  volume = {3}, number = {1}, pages = {1--26}, year = {2021},
  doi = {10.3934/fods.2021002}
}

@article{bhattacharya2026inferences,
  author = {Sohom Bhattacharya and Jianqing Fan and Jikai Hou},
  title = {Inferences on mixing probabilities and ranking in mixed-membership models},
  journal = {Journal of the American Statistical Association},
  pages = {1--12}, year = {2026}
}

@article{BradleyTerry1952,
  author = {Ralph Allan Bradley and Milton E. Terry},
  title = {Rank analysis of incomplete block designs: {I}. The method of paired comparisons},
  journal = {Biometrika},
  volume = {39}, number = {3/4}, pages = {324--345}, year = {1952},
  doi = {10.1093/biomet/39.3-4.324}
}

@article{ChatterjeeBhattacharya2026,
  author = {Sayak Chatterjee and Bhaswar B. Bhattacharya},
  title = {Transitivity in inhomogeneous random tournaments},
  journal = {arXiv preprint arXiv:2606.02340}, year = {2026}
}

@inproceedings{Chiang2024ChatbotArena,
  author = {Wei-Lin Chiang and Lianmin Zheng and Ying Sheng and Anastasios Nikolas Angelopoulos and Tianle Li and Dacheng Li and Banghua Zhu and Hao Zhang and Michael I. Jordan and Joseph E. Gonzalez and Ion Stoica},
  title = {{Chatbot Arena}: An open platform for evaluating {LLMs} by human preference},
  booktitle = {Proceedings of the 41st International Conference on Machine Learning},
  series = {Proceedings of Machine Learning Research},
  volume = {235}, pages = {8359--8388}, publisher = {PMLR}, year = {2024}
}

@article{Ford1957,
  author = {Ford, Jr., Lester R.},
  title = {Solution of a ranking problem from binary comparisons},
  journal = {The American Mathematical Monthly},
  volume = {64}, number = {8P2}, pages = {28--33}, year = {1957},
  doi = {10.1080/00029890.1957.11989117}
}

@article{GaoShenZhang2023,
  author = {Chao Gao and Yandi Shen and Anderson Y. Zhang},
  title = {Uncertainty quantification in the {Bradley--Terry--Luce} model},
  journal = {Information and Inference: A Journal of the IMA},
  volume = {12}, number = {2}, pages = {1073--1140}, year = {2023},
  doi = {10.1093/imaiai/iaac032}
}

@article{HemerikGoeman2018,
  author = {Jesse Hemerik and Jelle Goeman},
  title = {Exact testing with random permutations},
  journal = {TEST}, volume = {27}, number = {4}, pages = {811--825}, year = {2018},
  doi = {10.1007/s11749-017-0571-1}
}

@book{hoffman2022methods,
  author = {Mark Anthony Hoffman},
  title = {Methods for Network Analysis},
  publisher = {Bookdown}, year = {2022},
  url = {https://bookdown.org}
}

@article{Holm1979,
  author = {Sture Holm},
  title = {A simple sequentially rejective multiple test procedure},
  journal = {Scandinavian Journal of Statistics},
  volume = {6}, number = {2}, pages = {65--70}, year = {1979},
  url = {https://www.jstor.org/stable/4615733}
}

@inproceedings{huang2026dropping,
  author = {Jenny Huang and Yunyi Shen and Dennis Wei and Tamara Broderick},
  title = {Dropping just a handful of preferences can change top large language model rankings},
  booktitle = {International Conference on Learning Representations},
  volume = {2026}, pages = {90685--90721}, year = {2026}
}

@article{Jiang2011,
  author = {Xiaoye Jiang and Lek-Heng Lim and Yuan Yao and Yinyu Ye},
  title = {Statistical ranking and combinatorial {Hodge} theory},
  journal = {Mathematical Programming},
  volume = {127}, number = {1}, pages = {203--244}, year = {2011},
  doi = {10.1007/s10107-010-0419-x}
}

@article{KendallSmith1940,
  author = {Maurice G. Kendall and {Babington Smith}, B.},
  title = {On the method of paired comparisons},
  journal = {Biometrika},
  volume = {31}, number = {3--4}, pages = {324--345}, year = {1940},
  doi = {10.1093/biomet/31.3-4.324}
}

@article{Liu2008,
  author = {Yufeng Liu and D. Neil Hayes and Andrew Nobel and J. S. Marron},
  title = {Statistical significance of clustering for high-dimension, low-sample size data},
  journal = {Journal of the American Statistical Association},
  volume = {103}, number = {483}, pages = {1281--1293}, year = {2008},
  doi = {10.1198/016214508000000454}
}

@book{luce1959individual,
  author = {R. Duncan Luce},
  title = {Individual Choice Behavior: A Theoretical Analysis},
  publisher = {John Wiley \& Sons}, address = {New York}, year = {1959}
}

@article{makur2025minimax,
  author = {Anuran Makur and Japneet Singh},
  title = {Minimax hypothesis testing for the {Bradley--Terry--Luce} model},
  journal = {IEEE Transactions on Information Theory}, year = {2025}
}

@article{oliveira2024hodgerank,
  author = {Luna RN Oliveira and Jos{\'e} T Lunardi and Marcos Cal{\c{c}}ada and Ana L Pereira and De Jesuz, Danilo AF and Cristina Costa},
  title = {{HodgeRank} as a new tool to explore the structure of a social representation},
  journal = {Frontiers in Physics}, volume = {12}, pages = {1333727}, year = {2024}
}

@article{PhipsonSmyth2010,
  author = {Belinda Phipson and Gordon K. Smyth},
  title = {Permutation p-values should never be zero: Calculating exact p-values when permutations are randomly drawn},
  journal = {Statistical Applications in Genetics and Molecular Biology},
  volume = {9}, number = {1}, pages = {Article 39}, year = {2010},
  doi = {10.2202/1544-6115.1585}
}

@article{regenwetter2010testing,
  author = {Michel Regenwetter and Jason Dana and Clintin P Davis-Stober},
  title = {Testing transitivity of preferences on two-alternative forced choice data},
  journal = {Frontiers in Psychology}, volume = {1}, pages = {148}, year = {2010}
}

@article{ShahWainwright2018,
  author = {Nihar B. Shah and Martin J. Wainwright},
  title = {Simple, robust and optimal ranking from pairwise comparisons},
  journal = {Journal of Machine Learning Research},
  volume = {18}, number = {199}, pages = {1--38}, year = {2018},
  url = {https://jmlr.org/papers/v18/16-206.html}
}

@article{ShahEtAl2016,
  author = {Nihar B. Shah and Sivaraman Balakrishnan and Joseph Bradley and Abhay Parekh and Kannan Ramchandran and Martin J. Wainwright},
  title = {Estimation from pairwise comparisons: Sharp minimax bounds with topology dependence},
  journal = {Journal of Machine Learning Research},
  volume = {17}, number = {58}, pages = {1--47}, year = {2016},
  url = {https://jmlr.org/papers/v17/15-189.html}
}

@article{SinghDavidov2024,
  author = {Rahul Singh and Ori Davidov},
  title = {Modeling cyclicality and intransitivity in paired comparisons data},
  journal = {arXiv preprint arXiv:2406.11584}, year = {2024}
}

@article{SinghDavidov2026,
  author = {Rahul Singh and Ori Davidov},
  title = {Testing for lack of fit in paired comparison data},
  journal = {arXiv preprint arXiv:2604.00426}, year = {2026}
}

@article{SinghIliopoulosDavidov2025,
  author = {Rahul Singh and George Iliopoulos and Ori Davidov},
  title = {Least squares for cardinal paired comparisons data},
  journal = {Journal of the Royal Statistical Society: Series B (Statistical Methodology)},
  volume = {87}, number = {5}, pages = {1678--1706}, year = {2025},
  doi = {10.1093/jrsssb/qkaf035}
}

@article{thurstone1927law,
  author = {Louis L. Thurstone},
  title = {A law of comparative judgment},
  journal = {Psychological Review},
  volume = {34}, number = {4}, pages = {273--286}, year = {1927},
  doi = {10.1037/h0070288}
}

@article{UematsuLee2015,
  author = {Kazuki Uematsu and Yoonkyung Lee},
  title = {Statistical optimality in multipartite ranking and ordinal regression},
  journal = {IEEE Transactions on Pattern Analysis and Machine Intelligence},
  volume = {37}, number = {5}, pages = {1080--1094}, year = {2015},
  doi = {10.1109/TPAMI.2014.2360397}
}

@article{UematsuLee2017,
  author = {Kazuki Uematsu and Yoonkyung Lee},
  title = {On theoretically optimal ranking functions in bipartite ranking},
  journal = {Journal of the American Statistical Association},
  volume = {112}, number = {519}, pages = {1311--1322}, year = {2017},
  doi = {10.1080/01621459.2016.1208829}
}

@article{Wilcoxon1945,
  author = {Frank Wilcoxon},
  title = {Individual comparisons by ranking methods},
  journal = {Biometrics Bulletin},
  volume = {1}, number = {6}, pages = {80--83}, year = {1945},
  doi = {10.2307/3001968}
}
\end{document}